\documentclass{article}
\usepackage{amssymb}

\usepackage{graphicx}
\usepackage{amsmath}

\input{tcilatex}

\begin{document}

\title{\textbf{Strangelets and their possible astrophysical origin}\thanks{
Talk given at PASI on New States of Matter in Hadronic Interactions, Campos
do Jord\~{a}o 7-18 January 2002}}
\author{Luis Masperi\thanks{
On leave of absence from Centro At\'{o}mico Bariloche, 8400 San Carlos de
Bariloche, Argentina.} \\
Centro Latinoamericano de F\'{i}sica, Av. Venceslau Br\'{a}z 71 Fundos\\
22290-140 Rio de Janeiro, Brazil}
\date{}
\maketitle

\begin{abstract}
We discuss the first-order phase transition of QCD at high temperature in
the universe and the possible formation of quark-matter lumps through
cooling in regions of increased pressure. We show the similarity of results
using confinement and breaking of chiral symmetry. The inclusion of strange
matter may give stability to small drops, or strangelets, in the
colour-flavour locked phase possibly achieved in neutron stars that could
reach high-altitude observatory like that of Chacaltaya as primary cosmic
rays.
\end{abstract}

\section{ Introduction}

The phase diagram of quantum chromo dynamics $(QCD)$ is receiving a
particular attention since it is becoming possible to achieve the
quark-gluon plasma $(QGP)$ through heavy ions collisions in accelerators at
Brookhaven - $RHIC$ and CERN-SPS, and in the next future at ALICE.

The equivalent transition in the reversed direction generating hadrons in
the primordial universe with the possible formation of nuggets of very dense
quark-matter remnants has cosmological consequences like the baryonic matter
content.

Furthermore quark matter stabilized by a strange component may exist in
neutron stars which should be electrically neutral if pressure is high
enough to reach the colour-flawour locked phase $(CFL)$ with equal number of 
$u$, $d$ and $s$. The collision of two neutron stars might eject small
drops, strangelets, with a slight positive charge which could be absolutely
stable if their equivalent atomic number is larger than a minimum.

In Section 2 we describe the first-order confinement transition through the
generation of hadronic bubbles in the phase of independent quarks. When the
transition is almost completed, lumps of quark matter may be pressed and
survive with influence on the baryonic content necessary for primordial
nucleosyn-thesis. The equivalent mechanism with vacuum which breaks the
chiral symmetry is also considered.

In Section 3 the effect of stabilization due to the addition of a strange
component is discussed with particular emphasys on the phases of $QCD$
caused by the quark pairing at extremely high density.

Finally in Section 4 astrophysical conclusions are given about the
possibility that strangelets reach high-altitude observatories like the
Chacaltaya one and even evade the ordinary $GZK$ cutoff giving another
alternative for the explanation of ultra-high energy cosmic rays $(UHECR)$.

\section{ First-order phase transition of QCD}

Considering that in high-temperature universe above $T_{c}\sim 100$ $MeV$\
we have non-interacting relativistic quarks, the grand-canonical partition
function for fermions will give their pressure and density

\begin{equation}
p\simeq \frac{1}{6\pi ^{2}}\int_{0}^{\infty }dk\text{ }k^{3}\text{ }\frac{1}{%
e^{\frac{1}{T}\left( k-\mu \right) }+1}\text{ },  \label{1a}
\end{equation}

\begin{equation}
\frac{N}{V}\simeq \frac{1}{2\pi ^{2}}\int_{0}^{\infty }dk\text{ }k^{2}\text{ 
}\frac{1}{e^{\frac{1}{T}\left( k-\mu \right) }+1}\text{ }.  \label{1b}
\end{equation}

For chemical potential $\mu <T$ , the quark (antiquark) densities and
pressures can be appreximated by

\begin{equation}
\frac{N_{q\left( \overline{q}\right) }}{V}\simeq \frac{2.7}{\pi ^{2}}\text{ }%
T^{3}\pm \frac{1.6}{\pi ^{2}}\text{ }T^{2}\text{ }\mu _{q}\text{ },
\label{2a}
\end{equation}

\begin{equation}
p_{q\left( \overline{q}\right) }\simeq \frac{3.8}{\pi ^{2}}\text{ }T^{4}\pm 
\frac{2.7}{\pi ^{2}}\text{ }T^{3}\text{ }\mu _{q}\text{ },  \label{2b}
\end{equation}
so that the baryon number density and total pressure are

\begin{equation}
\frac{N_{q}-N_{\overline{q}}}{V}=n_{B}\simeq \frac{3.2}{\pi ^{2}}\text{ }%
T^{2}\text{ }\mu _{q}\text{ },  \label{3a}
\end{equation}

\begin{equation}
p_{q}+p_{\overline{q}}=p\simeq \frac{7.6}{\pi ^{2}}\text{ }T^{4}\text{ }.
\label{3b}
\end{equation}

Therefore since the entropy density $s\sim T^{3}$ the normalized
matter-antimatter asymmetry is

\begin{equation}
\QOVERD. ) {n_{B}}{s}^{HT}\sim \frac{\mu _{q}}{T}\sim 10^{-10}\text{ },
\label{4}
\end{equation}
from which $\mu _{q}\sim 10^{-2}$ $eV$ at the transition temperature.

During the transition, when $T$\ is constant, the chemical potential is
equal in both phases and in the confined one where $m_{N}\simeq 940$ $MeV>T$
it turns out that

\begin{equation}
\QOVERD. ) {n_{B}}{s}^{LT}\sim \left( \frac{m_{N}}{T}\right) ^{\frac{3}{2}}%
\text{ }e^{-\frac{m_{N}}{T}}\text{ }\frac{\mu _{q}}{T}=\varepsilon \text{ }%
\QOVERD. ) {n_{B}}{s}^{HT}\text{ },  \label{5}
\end{equation}
with $\varepsilon \sim 10^{-2}$\ , so that the matter-antimatter asymetry
becomes diluted by the increase of the total entropy.

But the percolation process of hadronic bubbles ends when the universe
expansion prevails on coalescence which occurs\cite{E. Witten} for a radius
of them

\begin{equation}
R_{1}\sim \frac{1}{T_{c}}\left( \frac{m_{pl}}{T_{c}}\right) ^{\frac{2}{3}%
}\sim 1\text{ }cm\text{ },  \label{6}
\end{equation}
and afterwards a new stage of quark-matter bubbles starts.

If these lumps of high-temperature phase do not evaporate but keep the
initial baryonic number $N_{B}\sim 10^{30}$\ while quark-antiquark pairs
annihilate, the free energy of each bubble will be

\begin{equation}
G\simeq R^{3}\text{ }T^{4}+R^{3}\text{ }T^{2}\text{ }\mu ^{2}\simeq R^{3}%
\text{ }T^{4}+\frac{N_{B}^{2}}{T^{2}\text{ }R^{3}}\text{ .}  \label{7}
\end{equation}

Since initially, for $R\simeq R_{1}$\ , $\mu <<T$\ the first term dominates
and the bubble contracts, but afterwards $\mu $\ inside increases and is no
longer equal to the value outside. Therefore, if the process is fast enough
to consider $T$\ constant, the contraction will stop when

\begin{equation}
\frac{\partial G}{\partial V}=0\simeq T^{4}-\frac{N_{B}^{2}}{T^{2}}\text{ }%
\frac{1}{V^{2}}  \label{8}
\end{equation}
so that $\overline{R}\simeq \frac{N_{B}^{\frac{1}{3}}}{T}\simeq 10^{-3}$ $cm$%
\ and $\mu \sim T$ , $i.e.$ we are reaching the limit of validity of Eqs.(3,
4).

For $\mu >T$ and when only quarks remain Eqs.(1, 2) can be approximated by

\begin{equation}
n_{B}\simeq \frac{\mu _{q}^{3}}{6\pi ^{2}}\text{ \ \ \ , \ \ \ }p\simeq 
\frac{\mu _{q}^{4}}{24\pi ^{2}}\text{ }.  \label{9}
\end{equation}

It is easy to see that the decrease of volume for the annihilation of all
the antiquarks in terms of the initial volume $\left( R\simeq R_{1}\right) $
and chemical potential (equal inside and outside the bubble) is

\begin{equation}
\Delta V\simeq V^{in}\text{ }\left( 1-\frac{3.2}{5.4}\text{ }\frac{\mu
_{q}^{in}}{T}\right) \text{ },  \label{10}
\end{equation}
so that the volume for the beginning of the case of Eq.(12) is

\begin{equation}
\overline{V}\sim \frac{\mu _{q}^{in}}{T}\text{ }V^{in}\simeq 10^{-10}\text{ }%
V^{in}\text{ },  \label{11}
\end{equation}
corresponding to $\overline{R}\sim 10^{-3}$ $cm$\ .This confirms that the
stabilization of the bubble occurs in the intermediate region of validity
between the two cases $\mu \lessgtr T$.

One thinks that confinement and breaking of chiral symmetry are related
phenomena. The phase transition for the latter is understood as a change of
vacuum with a corresponding decrease of energy $\Delta U_{vac}\left(
T\right) $ with the limits $\Delta U_{vac}\left( T_{c}\right) =0$ and $%
\Delta U_{vac}\left( 0\right) \simeq \Lambda _{QCD}^{4}$ . If one considers
a bubble of high-temperature vacuum, it contracts like a fluid so that its
free energy will be

\begin{equation}
G\simeq \Delta U_{vac}\text{ }R^{3}+p\left( R_{1}^{3}-R^{3}\right) \text{ },
\label{12}
\end{equation}
with $pR^{3}\sim T$\ . The contraction will stop for

\begin{equation}
\overline{R}\simeq \left( \frac{R_{1}^{3}\text{ }T}{\Delta U_{vac}}\right) ^{%
\frac{1}{6}}\sim 10^{-3}\text{ }cm\text{ },  \label{13}
\end{equation}
if $\Delta U_{vac}\sim $ $T^{2}$ $\mu _{q}^{in\text{ }2}$\ . It is clear
that with the vacuum interpretation the chemical potential has no physical
meaning and is introduced for normalization.

One must remark that the chemical potential had to be zero in the epoch of
baryon number non-conservation which generated the matter-antimatter
asymmetry. The value $\mu _{q}\sim 10^{-2}$ $eV$\ for quarks near the
confinement transition is what is needed for $\frac{n_{B}}{s}$\ to give the
primordial nucleosynthesis. But after the phase transition most of baryonic
number is kept inside the bubbles which are not expected to remain stable
for decreasing temperature. Since these lumps do not participate to
nucleosynthesis, their lifetime may affect what is normally predicted from
the evidence of this process regarding cosmology.

\section{ Stabilized strange quark matter}

\qquad It is possible to see that in the approximation of negligible mass of
quark $s$\ , strange quark matter may be more stable than that of only
almost massless $u$\ and $d$\ . In fact for the latter case the quark number
densities must be

\begin{equation}
\frac{N_{u}}{V}\sim \mu ^{3}\text{ \ \ \ , \ \ \ }\frac{N_{d}}{V}\sim 2\mu
^{3}\text{ },  \label{14}
\end{equation}
if we wish to have neutral matter. Therefore the pressure will be

\begin{equation}
p\sim \mu ^{4}+2^{\frac{4}{3}}\text{ }\mu ^{4}\text{ }.  \label{15}
\end{equation}

With the addition of the strange quark component, equal number densities of
quarks $u$, $d$\ and $s$\ will give neutral matter and a common chemical
potential $\widetilde{\mu }$\ in terms of which

\begin{equation}
p\sim 3\widetilde{\mu }^{4}\text{ }.  \label{16}
\end{equation}

With the condition of equal pressure in both cases Eqs.(18, 19) the ratio of
kinetic energy of $1$ $u$ , $1$ $d$ , $1$ $s$ and $1$ $u$ , $2$ $d$\ is

\begin{equation}
\frac{3\widetilde{\mu }}{\mu +2\text{ }2^{\frac{1}{3}\text{ }}\mu }\simeq
0.89\text{ },  \label{17}
\end{equation}
giving more stability to strange quark-matter.

Intuitively surface effects give a positive charge to lumps of strange
quark-matter\cite{Madsen}. This is because the larger mass of $s$\ makes the
corresponding wave function vanish on the surface decreasing its number
there.

At low temperature and very high pressure $QCD$ has a superconducting phase 
\cite{K. Rajagopal} which could lead to more stable strangelets. The
appearance of this new state of matter may be understood because due to
asymptotic freedom the distribution of quarks, which depends on $E-N\mu $\ ,
does not change if one is added on Fermi surface increasing the total energy 
$E$\ in $\mu $\ and the number in $1$\ . But including the attractive
exchange of gluon between a pair produces a bound $E_{B}$\ which gives a
corresponding decrease of $E-N\mu $\ favouring the condensation and a
fermionic gap.

There are two situations\cite{R. Casalbuoni} according with the number of
flawours which are taken as massless.

With two light quarks $u$\ and $d$\ , the condensate is

\begin{equation}
<q_{i\text{ }L\left( R\right) }^{\alpha }\text{ }C\text{ }q_{j\text{ }%
L\left( R\right) }^{\beta }>\text{ }\sim \Delta \text{ }\varepsilon _{ij}%
\text{ }\varepsilon ^{\alpha \beta \text{ }3}\text{ },  \label{18}
\end{equation}
with colour indices $\alpha $ , $\beta =1,$ $2,$ $3$\ and flawour ones $i,$ $%
j=1,$ $2$\ , being the charge conjugation matrix $C=i$ $\gamma ^{2}$ $\gamma
^{0}$\ . This colour superconductivity of $2$\ quarks $(2SC)$ corresponds to
a symmetry breaking

\begin{equation*}
SU\left( 3\right) _{c}\times SU\left( 2\right) _{L}\times SU\left( 2\right)
_{R}\longrightarrow SU\left( 2\right) _{c}\times SU\left( 2\right)
_{L}\times SU\left( 2\right) _{R}
\end{equation*}

The chiral global symmetrry $SU(2)$ remains unbroken, the fermionic gap
affects colours $1$ and $2$ and five gluons become massive.

The other possible condensate occurs for three light quarks $u$, $d$ and $s$ 
$(i,j=1,2,3)$

\begin{equation*}
<q_{i\text{ }L\left( R\right) }^{\alpha }\text{ }C\text{ }q_{j\text{ }%
L\left( R\right) }^{\beta }>\text{ }\sim \Delta \text{ }\delta _{i}^{\alpha }%
\text{ }\delta _{j}^{\beta }+\Delta ^{\prime }\text{ }\delta _{j}^{\alpha }%
\text{ }\delta _{i}^{\beta }
\end{equation*}
corresponding to the breaking

\begin{equation*}
SU\left( 3\right) _{c}\times SU\left( 3\right) _{L}\times SU\left( 3\right)
_{R}\times U\left( 1\right) _{B}\times U\left( 1\right) _{A}\longrightarrow
SU\left( 3\right) _{c+L+R}\text{ }\times Z_{2}\times Z_{2}
\end{equation*}

In this colour-flawour locking $(CFL)$ the $8$ gluons become massive and all
the fermions are gapped. The global chiral $SU(3)$ symmetry is broken, as
well as the baryonic $U(1)_{B}$ and the axial $U(1)_{A}$ giving therefore a
total of $8+2=10$ Goldstone bosons.

This latter condensate is more suitable for stable strangelets\cite{J.
Madsen}.

In fact this bulk phase is electrically neutral because of the equal number
of $u$, $d$ and $s$, and more stable than ordinary quark matter due to the
pairing. The surface effect decreases the $s$ contribution and makes the $%
CFL $ strangelet certainly positive avoiding catastrophic possibilities
analyzed in general context\cite{A. Dar}. This strangelet may be absolutely
stable, $i.e.$ having energy per baryon $<938$ $MeV$ , for atomic number $%
A>A_{\min }\sim 300$\ . The relation with charge is $Z$ $\propto A^{\frac{2}{%
3}}$\ because $Z$ depends on surface whereas $A$ comes from all the volume.
This law is different from that of ordinary strangelets which is $Z$ $%
\propto $ $A $\ for small ones and $Z$ $\propto $ $A^{\frac{1}{3}}$\ for $%
A\gtrsim 150$\ .

\section{ Astrophysical conclusions}

\qquad Presumably the strangelets formed at the beginning of universe were
ordinary and decayed by weak interaction emitting neutrons.

On the other hand, $CFL$ phase could be formed in neutron stars and from
collision of two of them the corresponding strangelets might be emitted and
reach our atmosphere. Their initial $A\sim 10^{3}$\ may decrease through
collisions and finally they should disintegrate when becomes $<$ $A_{\min }$%
\ .

There are events in cosmic rays which can correspond to strangelets: $2$
with \cite{T. Saito} $Z\simeq 14$\ and $A\simeq 350$ , $450$\ ; $1$ with\cite
{P.B. Price} $Z\simeq 46$\ and $A>1000$\ ; $1$ with\cite{M. Ichimura} $Z=20$%
\ and $A\simeq 460$ and the Centauro\cite{C. Lattes} of $A\sim 200$\ .

Strangelets, due to their large $A$, propagate in $CMB$ with small loss of
energy\cite{M. Rybazynski} and have a $GZK$ cutoff $\sim 10^{21}$ $eV$\ ,
larger than that for protons, so that they might explain the $UHECR$.

If their initial atomic number is $\sim 2000$\ they could arrive to a
high-altitude observatory, like the Chacaltaya\cite{O. Saavedra} one, with $%
A>A_{\min }$\ and be detected directly apart from the $EAS$ which would have
an energy spectrum of secondary hadrons softer than that caused by primary
protons.

The possible detection of strangelets is also planned\cite{J.Madsen} with
the AMS-02 on the International Space Station for 2005.

\bigskip 
\begin{equation*}
\end{equation*}

\begin{equation*}
\text{{\Large Acknowledgments}}
\end{equation*}
Thanks are due to Francesco Calogero and Oscar Saavedra for extremely
stimulatings comments regarding theoretical and experimental aspects of
strangelets, and to Erasmo Ferreira and all the organizers of the Pan
American meeting.\qquad

\end{document}